\documentclass{biometrika}

\usepackage[T1]{fontenc}
\usepackage{amsmath}
\usepackage{amsfonts}
\usepackage{amssymb}
\usepackage{times}
\usepackage{graphicx,color}
\usepackage{caption}
\usepackage{subcaption}
\usepackage{natbib}
\usepackage{verbatim}

\usepackage{bm}
\usepackage[plain,noend]{algorithm2e}

\makeatletter
\renewcommand{\algocf@captiontext}[2]{#1\algocf@typo. \AlCapFnt{}#2} 
\def\@algocf@capt@plain{top}
\renewcommand{\algocf@makecaption}[2]{%
  \addtolength{\hsize}{\algomargin}%
  \sbox\@tempboxa{\algocf@captiontext{#1}{#2}}%
  \ifdim\wd\@tempboxa >\hsize
    \hskip .5\algomargin%
    \parbox[t]{\hsize}{\algocf@captiontext{#1}{#2}}
  \else%
    \global\@minipagefalse%
    \hbox to\hsize{\box\@tempboxa}
  \fi%
  \addtolength{\hsize}{-\algomargin}%
}
\makeatother

\usepackage{willcommands}

\renewcommand{\Var}{\operatorname{var}}
\renewcommand{\Cov}{\operatorname{cov}}

\def\spacingset#1{\renewcommand{\baselinestretch}%
{#1}\small\normalsize} \spacingset{1}

\newcommand{\heta}{\hat{\eta}}
\newcommand{\hmu}{\hat{\mu}}

\newcommand{\hp}{\hat{p}}
\newcommand{\EE}{E}
\newcommand{\Bias}{\operatorname{bias}}
\newcommand{\MSE}{\operatorname{MSE}}
\newcommand{\Corr}{\operatorname{corr}}

\newcommand{\PP}{\operatorname{pr}}
\newcommand{\tX}{\widetilde{X}}
\newcommand{\tY}{\widetilde{Y}}
\newcommand{\oo}{\mathcal{O}}

\newcommand{\hF}{\widehat{F}}
\newcommand{\hG}{\widehat{G}}

\newcommand{\aVar}{\widetilde{\Var}}
\newcommand{\aBias}{\widetilde{\Bias}}
\newcommand{\aMSE}{\widetilde{\MSE}}
\newcommand{\aEE}{\widetilde{\EE}}

\newcommand{\htGH}{\hat{t}^\text{GH}}
\newcommand{\hkGH}{\hat{k}^\text{GH}}

\renewcommand{\gv}{\mid}
\renewcommand{\1}{1}

\begin{document}

\jname{}
\jyear{}
\jvol{}
\jnum{}
\copyrightinfo{}


\markboth{W. Fithian and S. Wager}{Semiparametric Exponential Families}

\title{Semiparametric exponential families for heavy-tailed data}

\author{WILLIAM FITHIAN \and STEFAN WAGER}
\affil{Department of Statistics, Stanford University, Stanford, California 94305, U.S.A.
\email{wfithian@stanford.edu} \email{swager@stanford.edu}}

\maketitle

\begin{abstract}
We propose a semiparametric method for fitting the tail of a heavy-tailed population given a relatively small sample from that population and a larger sample from a related background population. We model the tail of the small sample as an exponential tilt of the better-observed large-sample tail, using a robust sufficient statistic motivated by extreme value theory. In particular, our method induces an estimator of the small-population mean, and we give theoretical and empirical evidence that this estimator outperforms methods that do not use the background sample. We demonstrate substantial efficiency gains over competing methods in simulation and on data from a large controlled experiment conducted by Facebook.
\end{abstract}


\section{Introduction}

We study estimation of the tail of a distribution given a medium-sized sample from a heavy-tailed population of interest
$X_i \sim F(x)$, $i=1, \, \ldots, \, n,$
and a much larger background sample from a qualitatively similar but non-identical population
$ Y_i \sim F_0(x)$, $i=1, \, \ldots, \, N \gg n,$
with all $n+N$ observations independent.
As a motivating example, consider an internet company with millions of users that wants to apply some treatment, a change to the site, to a random subset of users and estimate the effect of the treatment on revenue per user.
Our goal is to understand the distribution of the smaller treatment sample, while customers who were not part of the experiment act as a background sample.
In such applications, the distribution of revenue is usually heavy-tailed. If, say, 10\% of the company's revenue comes from the top 0.1\% of its users, then understanding the tail of $F$ is crucial when estimating how much revenue will be gained or lost if the treatment is extended to all users.

Estimating the tail of $F$ from the $X$ sample alone is difficult because it will only contain a few extreme values. Approaches that only use $X_1, \, \ldots, \, X_n$ either suffer from high variability, or require strong parametric assumptions that can lead to large bias. We will use the background dataset to navigate this trade-off, and to produce stable yet accurate estimates of the tail of $F$.

Our key assumption is that the tails of $F$ and $F_0$ are similar enough that we can model the tail of $F$ as a perturbation of the tail of $F_0$. If so, we can translate stable estimates of the tail of $F_0$ into good estimates of the tail of $F$. In the spirit of \citet{efron1996using}, we model the tail of $F$ as an exponential tilt of the tail of $F_0$.  For some threshold $t$, a tuning parameter, define the conditional right tail law $G(x)=\PP(X-t\leq x \gv X>t)$, and $G_0$
analogously.  We model $G$ as an exponential family with carrier measure $G_0$:
 $ dG(x) = e^{\eta T(x) - \psi(\eta)}dG_0(x)$.
The performance of any such approach depends on the choice of sufficient statistic $T(x)$. By exploiting results from extreme value theory, we derive a sufficient statistic tailored to the tails of heavy-tailed distributions. Generalizing our method to estimate a heavy left tail is straightforward.

Our semiparametric framework is closely related to density ratio models \citep{fokianos2001semiparametric,fokianos2004merging,huang2012proportional,tan2009note}, which are usually fitted by empirical likelihood methods \citep{owen2001empirical}. In particular, \citet{de2013spectral} use a density ratio model to estimate a family of spectral densities of multivariate extreme value distributions indexed by covariates. Our sufficient statistic could also be used for density ratio modeling with heavy-tailed data.

Our method gives a generic approach to estimating $F$, with an emphasis on its tail. While we focus our analysis on the behavior of the mean estimator $\hmu$ induced by $\hF$, our estimator $\hF$ could also be used for other purposes such as large quantile estimation, semiparametric bootstrapping, or density estimation.

\section{Semiparametric tail estimation}
\label{secApproach}

\subsection{Constructing an exponential family}
\label{secSemipar}

Our goal is to model the tail law $G$ of $X$. Direct approaches to fitting $G$ might specify a parametric model for it, and then estimate the relevant parameters using $X_1,\ldots,X_n$. Instead, we propose a semiparametric method that models $\lambda(x) = dG(x)/dG_0(x)$, assuming absolute continuity of $dG$ with respect to $dG_0$.
Using this approach we can specify a simple model for $\lambda$ while preserving idiosyncrasies of the carrier measure
$G_0$ such as clustering or rounding effects.
Using the relation $G(x)=\int_0^x \,dG(u) = \int_0^x\lambda(u)\,dG_0(u)$, we can turn an estimator of $\lambda(x)$ into an estimator of $G$ by summing over the background tail points, weighted by $\hat\lambda(Y_i-t)$:
\begin{equation*}
\hG(x) = \frac{1}{\sum_{Y_i > t} \hat\lambda(Y_i-t)}\sum_{Y_i > t} \hat\lambda(Y_i-t) \1_{\{Y_i - t \,\leq\, x\}}.
\end{equation*}
We use the model $\lambda(x)=\exp\{\eta \, T(x) - \psi(\eta)\}$, a log-linear family indexed by $\eta\in\R$. The family of candidate distributions for $G$ is thus an exponential family with carrier $G_0$, sufficient statistic $T$, and normalizing constant $\exp\{\psi(\eta)\}$. The sufficient statistic $T$ controls the behavior and stability of the method, since maximum likelihood estimation in exponential families operates by moment matching on $T$. A good sufficient statistic should capture relevant information about the tail while remaining robust to the presence of very large observations.  For example, the identity map $T(x)=x$ used by, e.g., \citet{efron1996using} would not be a good choice for us, because a few very large $X$ values could dominate the sufficient statistic.

\subsection{Extreme value theory and the sufficient statistic}
\label{secSuffStat}

Extreme value theory provides a flexible and powerful framework for modeling the tails of distributions. See, e.g., \citet{beirlant2006statistics}, \citet{dehaan2006extreme}, or \citet{resnick2007heavy} for a review, and \citet{beirlant2012overview} for recent developments. Our sufficient statistic is motivated by a classical result: if $F$ is a heavy-tailed distribution with a regularly varying tail, then there is a sequence $\sigma_t$ and a constant $\gamma > 0$ such that, as $t \rightarrow \infty$,
\begin{equation}\label{eq:gpdLimit}
 \PP\left(\frac{X - t}{\sigma_t} \leq x \gv X > t\right) \to H_{\gamma, \, 1}(x),\quad H_{\gamma, \, \sigma}(x)= 1 - \left(1+\frac{\gamma x}{\sigma}\right)^{-1/\gamma},
\end{equation}
where $H_{\gamma,\sigma}$ is called a generalized Pareto distribution with tail index $\gamma$ and scale parameter $\sigma > 0$. As $\gamma\to 0$, $H_{\gamma,\sigma}$ tends to an exponential distribution.

Suppose that both our distribution of interest $F$ and the background $F_0$ both have regularly varying tails with the same tail index $\gamma > 0$. If the threshold $t$ is large enough for \eqref{eq:gpdLimit} to apply, we should expect the tail laws $G$ and $G_0$ to be well-approximated by generalized Pareto distributions with the same tail index $\gamma$ but with potentially different scales $\sigma$ and $\sigma_0$. Modeling different but related distributions as having common $\gamma$ and only allowing scale and location parameters to vary is not unusual, see for example \citet{davison1990models} or \citet[\S 6]{coles2001introduction}.

If $G$ and $G_0$ were really generalized Pareto distributions with $\sigma$ close to $\sigma_0$, we would have
\begin{align}
\label{eqFirstOrder}
  \log\lambda(x) &= \log({\sigma_0}) - \log(\sigma)
  + \frac{1+\gamma}{\gamma}\log(1+\gamma x/\sigma)
  - \frac{1+\gamma}{\gamma}\log(1+\gamma x/{\sigma_0}) \\
  &= \eta \,  \frac{x}{{\sigma_0}/\gamma + x}
- \psi(\eta) + \oo\left\{(\sigma - \sigma_0)^2\right\},
\nonumber
\end{align}
where $\psi$ and $\eta$ only depend on $\sigma_0$ and $\gamma$. This bound holds uniformly in $x\geq 0$.
Thus, under extreme value theoretic conditions, a linear tilting function with a sufficient statistic of the form $T(x) = x/(\kappa + x)$ should closely replicate the true relative density $\lambda(x)$.

Given~\eqref{eq:gpdLimit}, we could also try fitting the tail of $F$ directly using an extreme value theoretic model as advocated by, e.g., \citet{peng2001estimating}.
This approach, however, is vulnerable to model misspecification \citep[e.g.,][]{suveges2010model}.
A parametric generalized Pareto distribution fit would, for example, ignore any discretization or grouping effects from $\hG$, possibly giving a misleading picture of $G$. By contrast, we model $G$ as a perturbation of $G_0$, with \eqref{eq:gpdLimit} only motivating the direction of the perturbation. Thus, $\hG$ will reflect local idiosyncrasies of the background $G_0$.

\subsection{Our method in practice}
\label{secMethod}

We have proposed fitting the tail law $G$ as an exponential tilt of the background tail $G_0$, with sufficient statistic $T(x) = x/(\kappa + x)$. Carrying out our proposal requires estimating the tilt parameter $\eta$, as well as choosing a  bandwidth $\kappa$ and a threshold $t$.

Concerning $\eta$, if we had access to the full background distribution $G_0$, then the maximum likelihood estimator for $\eta$ would solve the moment-matching condition
\begin{equation}
\label{eqMatch}
\frac{1}{\sum_{X_i>t} 1}\sum_{X_i>t} T(X_i - t)
= \frac{\int_t^{\infty} T(y - t) \, e^{\heta \, T(y - t)}\,dG_0(y - t)}
{\int_t^{\infty} e^{\heta \, T(y - t)}\,dG_0(y - t)}.
\end{equation}
In applications, $G_0$ is unknown, but, as shown by \citet{owen2007infinitely}, we can obtain accurate estimates of $\heta$ by logistic regression when the size $N$ of background sample is large. To do this, we first join the sample of interest and the background sample into a single dataset, assigning the former observations a label 1 and the latter ones a 0, and then perform a logistic regression on this dataset with an intercept and with $T(x - t)$ as the predictor. Then, as $N \rightarrow \infty$, \citet{owen2007infinitely} showed that the slope parameter of the logistic regression converges to the solution to \eqref{eqMatch}.  In our analysis, we assume that $N$ is large enough for the error in \eqref{eqMatch} to be negligible. This assumption is reasonable in our motivating internet applications, as we typically have access to an extremely large store of background data. The results of \citet{owen2007infinitely} require moment conditions on the features, but they hold here because we are regressing on the bounded feature $T(x-t)$.

Second, our discussion from \S \ref{secSuffStat} suggests that setting $\kappa = \sigma_0/\gamma$ should be a good choice. In our experiments, we found that the simple approach of fitting $\gamma$ using the Hill estimator \citep{hill1975simple} on the background dataset and $\sigma_0$ by maximum likelihood worked well. We obtained very similar results using $\kappa = t$, which is motivated by the asymptotic limit $\sigma_t/(t\gamma) \rightarrow 1$ for $\sigma_t$ in \eqref{eq:gpdLimit}.

Finally, we must choose a threshold $t$. The most direct method, when possible, is to determine empirically what threshold works best in previously observed instances of the same problem. For example, the internet company may perform hundreds of
experiments each day, learning over time which values of $t$ work well for different kinds of problems. In the absence of historical data, an alternative method is required. One option is to use off-the-shelf threshold selection rules such as the method of \citet{guillou2001diagnostic} originally intended to set the threshold for the Hill estimator; other threshold estimation procedures are discussed in \citet[][\S 4.7]{beirlant2006statistics}. The Guillou--Hall procedure returns a number of observations $\hkGH$ to be used for tail index estimation; this value can be translated into a threshold $\htGH$ given by the $\hkGH$-th largest observation. In \S \ref{secTheory}, we show that if our goal is to estimate the mean of the $X_i$, then $\htGH$ grows to infinity at the correct asymptotic rate. However, in our experiments, the Guillou--Hall rule often picks larger-than-optimal thresholds. The weakness of the Guillou--Hall rule is that when $F$ and $F_0$ are very close to each other, we could use a fairly small threshold $t$ without suffering unduly high bias, since $dG(x)/dG_0(x)$ might be close to its limit even if $G$ and $G_0$ are not close to theirs. However, $\htGH$ only uses $X_1,\ldots,X_n$, and so has no way of detecting this. Developing an adaptive threshold selection procedure that efficiently uses both  $X_1, \, \ldots, \, X_n$ and $Y_1, \, \ldots, \, Y_N$, possibly following ideas from \citet{wadsworth2012likelihood}, is an interesting avenue for further research.

\section{Asymptotic theory for mean estimation}
\label{secTheory}

Combining our semiparametric estimator $\hG$ with the empirical law below $t$ yields an estimator $\hF$ for $F$,  which we can use to obtain plugin estimators for functionals of $F$ such as quantiles or moments if they exist. In this section, we analyze the asymptotic behavior of the mean estimator $\hmu$ induced by $\hF$:
\begin{equation*}
  \hmu = \int x \,d\hF(x) =  \frac{1}{n} \sum_{X_i \leq t} X_i
  + \frac{1}{n} \, \frac{\sum_{X_i>t} 1}{\sum_{Y_i>t}
    e^{\hat\eta \, T(Y_i-t)}}\sum_{Y_i>t}
  e^{\hat\eta \, T(Y_i-t)} \;Y_i.
\end{equation*}
When estimating $\mu$, we always assume that $F$ and $F_0$ both have a shared tail index $0 < \gamma < 1$, which implies that they have finite means. Because our sufficient statistic $T(x) = x / (\kappa + x)$ is bounded, the limit of $\hF$ as the background size $N\to\infty$ also has a tail index $\gamma < 1$, so $\hmu$ is well-defined in the large-$N$ limit. To simplify our analysis, we focus on this limit and assume that $N$ is large enough that the errors in $\hG_0$ are negligible. Proofs are deferred to the supplementary material.

Throughout this section, we assume that $F$ has second-order regularly varying tails
\begin{equation}
\label{eqSecond}
1 - F(x) = Cx^{-1/\gamma} \, \left\{1 + Dx^{-\beta} + o\left(x^{-\beta}\right)\right\}, 
\end{equation}
for some $\gamma$ and $\beta>0$,
and that the background distribution $F_0$ also satisfies this condition with the same $\gamma$ but possibly different values of $C$, $D$ and $\beta$. This condition is discussed at length by \citet[][\S 2.3]{dehaan2006extreme}; the form in \eqref{eqSecond}
was introduced by \citet{hall1982simple}. Our results are in terms of asymptotic moments, i.e., the moments of the limiting Gaussian random variable, which we denote as $\aVar$ and $\aEE$. We analyze a version of Winsorization that caps observations at a given threshold $t$ rather than at a predetermined quantile: $\hmu_{W} = n^{-1} \sum_{i = 1}^n \min(t, X_i)$. This version is more directly comparable to our method with fixed $t$.

\begin{theorem}
\label{theo:var}
Suppose that $F$ and $F_0$ satisfy~\eqref{eqSecond} with the same $0 < \gamma < 1$, and that the background sample size $N$ grows faster than $n^{1/\min(1, \, 2 - 2\gamma)}$. Then our estimator is asymptotically normal for any threshold sequence satisfying $t(n) = o(n^\gamma)$ and has asymptotic variance
  \begin{align}
  \label{eqDelta}
n \, \aVar\left(\hmu\right) = &F(t) \, \Var\left(X\mid X \leq t\right)
+ \left\{1-F(t)\right\}\ \frac{\Cov^2\left(T, \, X \mid X > t\right)}{\Var\left(T \mid X > t\right)} \\
\notag
&\ \ \ \ + F(t)\left\{1-F(t)\right\}\ \left\{\EE\left(X \mid X > t\right)- \EE\left(X \mid X \leq t\right)\right\}^2.
  \end{align}
\end{theorem}

We can turn this result into a variance estimator by plugging in $\hF$ for $F$ in \eqref{eqDelta}. The delta method gives us more intuition about why our semiparametric estimator is more stable than the sample mean. Whenever the sample mean has finite variance,
\begin{equation*}
n \left\{\Var\left(\bar{X}\right) - \aVar\left(\hmu\right)\right\} = \left\{1-F(t)\right\} \, \left\{1 - \Corr^2\left(T, \, X\mid X > t\right)\right\} \, \Var\left(X\mid X > t\right).
\end{equation*}
Thus, our method achieves a favorable bias--variance tradeoff if $T$ captures information relevant to estimating $\mu$ without being too correlated to $X$ itself.

For optimally chosen threshold sequences, our method achieves a better rate of convergence than Winsorization in the range $0.5 < \gamma < 1$, where $X$ has a finite mean but infinite variance.

\begin{theorem}
\label{theo:conv}
Suppose that $F$ and $F_0$ satisfy~\eqref{eqSecond} with the same $0.5 < \gamma < 1$, and that the background sample size $N$ grows faster than $n^{1/(2 - 2\gamma)}$. Then,
\begin{align}
\label{eqRate}
\aEE\left\{\left(\hmu^*_{S} - \mu\right)^2\right\} = \oo\left( n^{\left(2\gamma - 2 - 2\gamma\beta_{\textnormal{min}}\right)/
  \left(1 + 2\gamma\beta_{\textnormal{min}}\right)}\right), \quad
\EE\left\{\left(\hmu^*_{W} - \mu\right)^2\right\} = \oo\left( n^{2\gamma - 2}\right).
\end{align}
Here, $\beta_{\textnormal{min}}=\min(\beta, \, \beta_0)$, $S$ stands for our method, $W$ stands for Winsorization, and $\hmu^*_{\cdot}$ denotes each estimator computed at its optimal threshold.
\end{theorem}

Finally, we show below that we can estimate the optimal threshold $t^*$ using the method of \citet{guillou2001diagnostic}, as discussed in \S \ref{secMethod}.
\begin{corollary}
\label{coro:GH}
Write $\htGH$ for the threshold obtained by applying the Guillou--Hall method to $X_1,\ldots,X_n$.
Under the conditions of Theorem \ref{theo:conv} and assuming that $\beta = \beta_0$, the adaptive threshold sequence $\htGH$ grows at the same rate as the optimal threshold sequence $t^*$.
\end{corollary}

\section{Examples and experiments}
\label{secExperiments}

\subsection{Methods under comparison}

In this section, we apply our method both to simulated data and to real data provided by Facebook. We focus on mean estimation, the problem that originally motivated our research. In our experiments our method comfortably outperforms the baselines, which cannot take advantage of the background sample. Overall, the results show that the background sample carries useful information that can considerably improve estimates of $\mu$ if properly exploited, and that our semiparametric estimator can achieve this goal given a good choice of $t$.

Our first baseline is a version of Winsorization with fixed $t$ as discussed in \S\ref{secTheory}. When we need to choose $t$ adaptively, we use the second-largest observation as recommended by \citet{rivest1994statistical}. Our second baseline fits a generalized Pareto model to the tail of $X_1,\ldots,X_n$ by maximum likelihood \citep{johansson2003estimating}; a related idea was studied by \citet{peng2001estimating}. We report results for the method of Johansson at its oracle threshold $t$, as well as for $t$ selected by the method of \citet{guillou2001diagnostic}. Results for other parametric methods were similar.

\subsection{Simulation example}
\label{secSimStudy}

\begin{table}
\def~{\hphantom{0}}
   \tbl{Bias, variance, and mean squared error for mean estimators, log-gamma simulation. Our method outperforms Winsorization by about 30\%.}{%
  \begin{tabular}{lccc}
    Method & Variance \;(s.e.) & Bias$^2$ \;(s.e.) & MSE \;(s.e.) \\[5pt]
Semiparametric \;(Oracle $t$) & ~30 \;(1)~~ & 16 \;(1) & ~46 \;(2)~~\\
Semiparametric \;(Guillou--Hall) & ~33 \;(2)~~ & 26 \;(2) & ~59 \;(2)~~\\[4pt] Winsorized \;(Oracle $t$) & ~50 \;(2)~~ & 14 \;(2) & ~64 \;(3)~~\\
Winsorized \;($k=1$) & ~78 \;(5)~~ & 12 \;(2) & ~90 \;(5)~~\\ [4pt]
Pareto Tail \;(Oracle $t$) & ~64 \;(3)~~ & ~9 \;(2) & ~74 \;(3)~~\\
Pareto Tail \;(Guillou--Hall) & 472 \;(269) & 17 \;(6) & 488 \;(273)
  \end{tabular}}
\label{tab:simResults}
\begin{tabnote}
  MSE, mean squared error; s.e., standard error. All numbers multiplied by $100$.
\end{tabnote}
\end{table}

We begin by testing our method on data simulated from the log-gamma family: we drew $n=1000$ and $N=10^6$ values from the model $\log X_i \sim \textrm{Gamma}(k=4,s=0.45)$ and $ \log Y_i \sim \textrm{Gamma}(k=3,s=0.45)$, where $k$ and $s$ are the shape and scale parameters. Log-gamma distributions have regularly varying tails with $\gamma = s$, but do not satisfy \eqref{eqSecond} for any $\beta>0$. The different shape parameters make the two means very different: $\EE (Y) = 6.0$ while $\EE (X) = 10.9$.

\begin{figure}
    \centering
    \vspace{-3mm}
  \includegraphics[width=.63\textwidth]{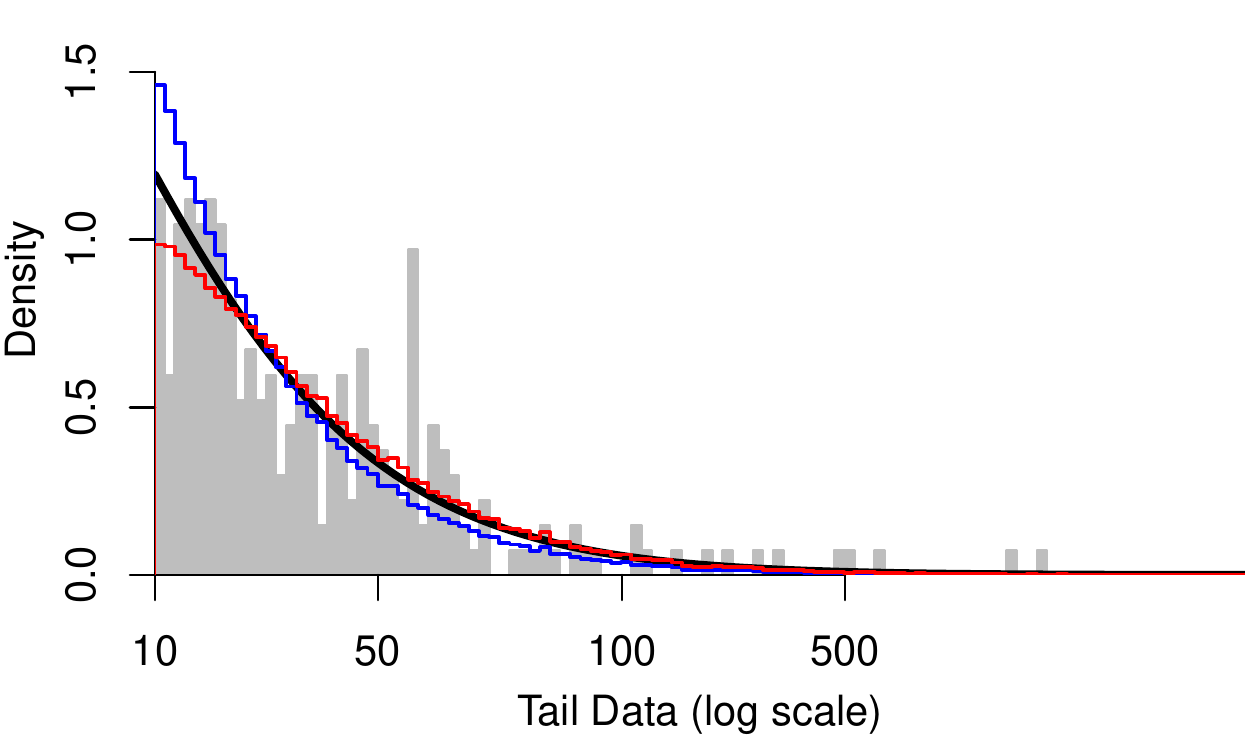}
  \vspace{-3mm}
   \caption{Illustration of our method on a single realization of the simulation of \S~\ref{secSimStudy}. Pictured are the true tail distribution $G$ of $X$ (solid black curve), the observed tail histograms of $X_i$ (solid gray histogram) and of $Y_i$ (outlined blue histogram), and the estimated tail law $\widehat G$ obtained by reweighting the $Y_i$ by $e^{\hat\eta T(Y_i-t)}$ (outlined red histogram).}
     \label{fig:fitTail}
\end{figure}

Table \ref{tab:simResults} shows results for all three methods, both at oracle and adaptive threshold choices. The oracle value is the value of $t$ minimizing the measured mean squared error. Even with an adaptive threshold choice, our method outperforms the oracle-$t$ baselines. The mean squared error for all the methods is much better than the variance of the unbiased sample mean, which is 9.9. Figure~\ref{figSimBiVarTO} shows the bias--variance tradeoff for both our method and Winsorization. We did not include the generalized Pareto distribution curve, as its behavior was erratic. The two methods are about equally biased at their respective oracle thresholds, but the semiparametric method has about 60\% of the variance of Winsorization. Our method's bias is negative in this case, perhaps because $\EE(Y)$ is much smaller than $\EE(X)$. In the supplementary material, we repeat this simulation for other values of $\gamma$ and test our method in the misspecified case where $X$ and $Y$ have different tail indices, with similar results throughout. Figure~\ref{fig:fitTail} illustrates our method on a single realization of the simulated data.

\subsection{Facebook data illustration}
\label{secFB}

\begin{figure}[t]
  \centering
\begin{subfigure}[b]{0.35\textwidth}
  \includegraphics[width=\textwidth]{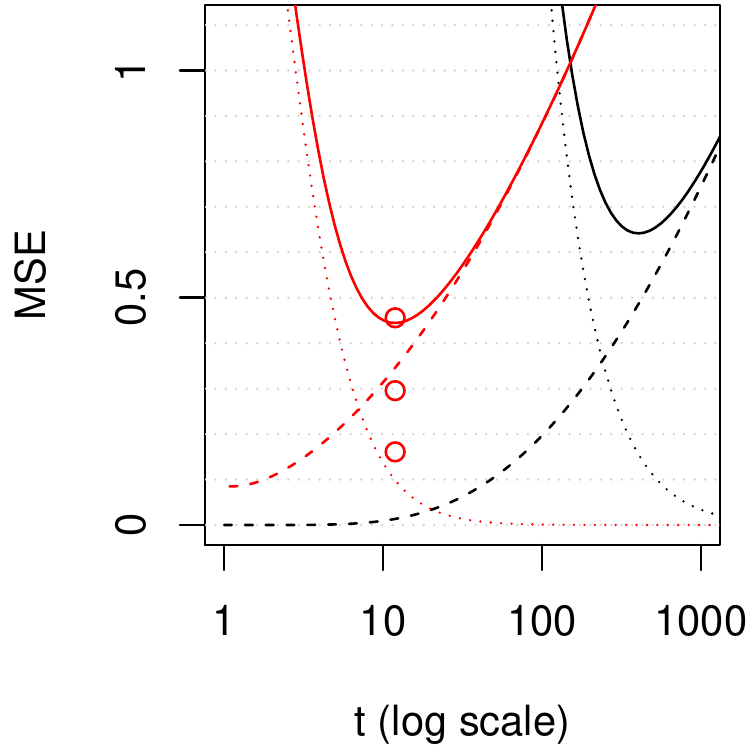}
\caption{Log-gamma experiment}
  \label{figSimBiVarTO}
\end{subfigure}
\begin{subfigure}[b]{0.58\textwidth}
  \includegraphics[trim = 10mm 0mm 0mm 0mm, clip=TRUE, width=\textwidth]{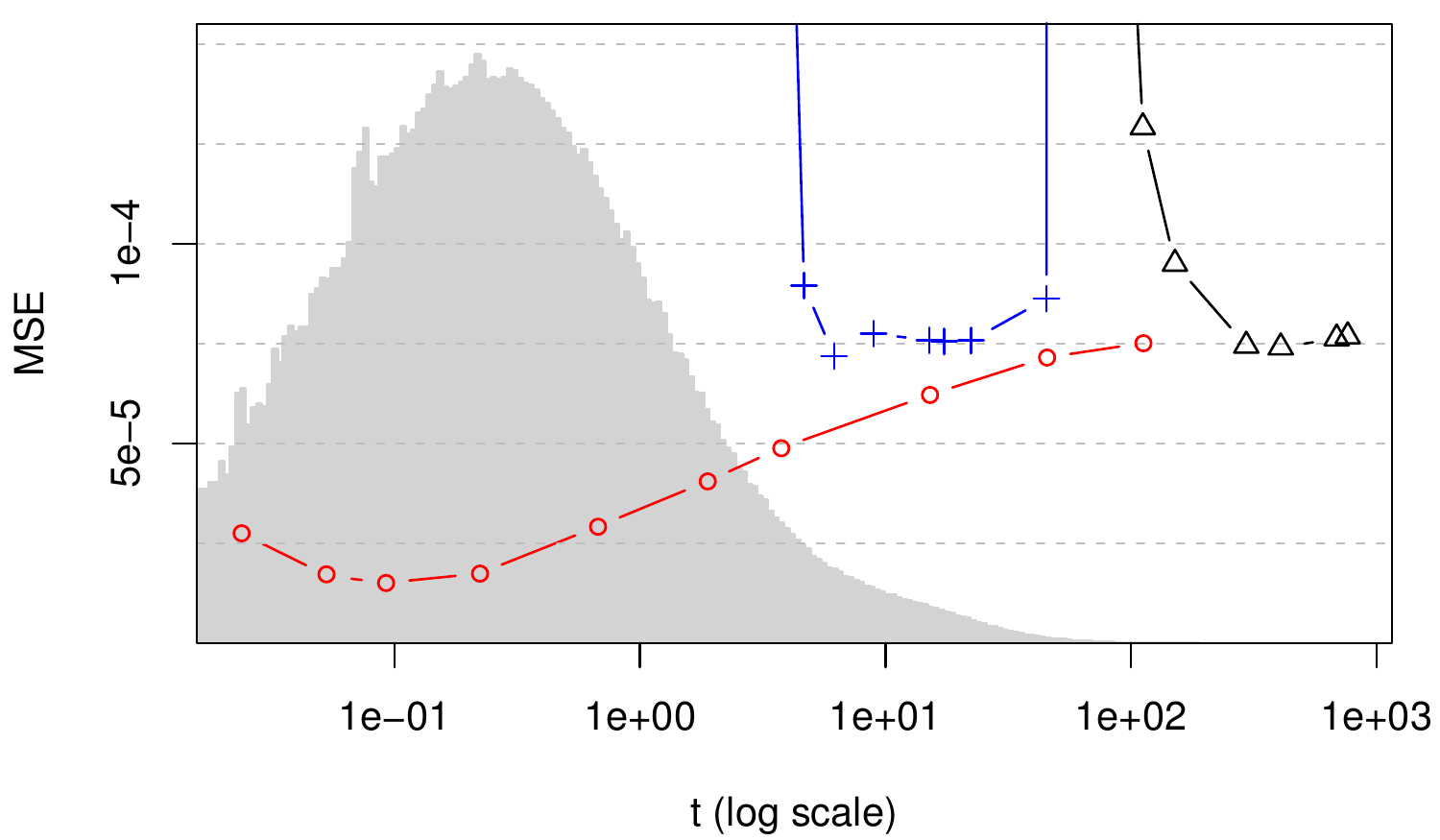}
\caption{Facebook experiment}
\label{fig:FB1}
\end{subfigure}
\vspace{-3mm}
\caption{Performance for mean estimation performance. (a) squared bias (dotted lines), variance (dashed), and mean squared error (MSE, solid) for mean estimation using Winsorization (black) and our method (red), in the simulation of \S 4.2, as a function of the threshold $t$. The plotted circles are Monte Carlo estimates for semiparametric variance, squared bias, and mean squared error at the optimal threshold of our method; we obtained the rest of the variance curve using \eqref{eqDelta} and the rest of the bias curve by setting $\kappa = t$ and computing bias on larger $X$-samples. (b) MSE for three mean estimates on the Facebook dataset: Winsorization (black, triangles), our method (red, circles), and parametric Pareto tail fitting (blue, crosses). A histogram of the background population is shown in solid gray.}
\end{figure}

Finally, we present results from applying our method to two arms of a large multi-arm experiment conducted by Facebook with about 5,000,000 observations each, representing advertising revenue for two different website layouts. To protect potentially sensitive information, we discarded users whose revenues were zero, then normalized the data so that the population of interest had unit mean.
For the purpose of our experiment, these two large samples comprise the population of interest $F$ and background population $F_0$, from which we sample smaller data sets with replacement.  This approach allows us to estimate bias and variance for each procedure. We apply our method to ${n = 200,000}$ observations drawn with replacement from $F$ and ${N = 3,000,000}$ from $F_0$. We evaluate each method by comparing its mean estimate with the sample mean of the 5,000,000 original data points. Figure \ref{fig:FB1} and Table \ref{tab:FB} show results using all three methods, averaged across 10,000 trials. For a wide range of thresholds $t$, our method outperforms its competitors at their own optimal thresholds.

The mean squared error numbers in Table \ref{tab:FB} may appear quite small at first. Recall, however, that our real goal in this experiment is to predict total annual advertising revenue if the new site layout is adopted. Supposing that the baseline annual revenue is around \$10 billion, then Winsorization with $k=1$ leads to a root mean squared error of $(80\times 10^{-6})^{1/2} \times \$10^9 = \$89$ million, whereas the error for our method using the 75th percentile $t$ is \$54 million. Thus, even with 200,000 users involved in the experiment, accurate revenue prediction is a difficult problem requiring statistically efficient methods. Our results suggest that, by using our method, Facebook could have made good use of the available background information to considerably improve their revenue estimates.

\section*{Acknowledgment}

We are grateful to Bradley Efron, Trevor Hastie, Robert Tibshirani and Guenther Walther for many helpful conversations, to Facebook for allowing us to report results on their dataset, and to the {\it Biometrika} editors and referees for providing constructive feedback and suggestions that greatly improved our paper. W. F. and S. W. are supported by an NSF VIGRE grant and by a B. C. and E. J. Eaves Stanford Graduate Fellowship respectively.


\begin{table}[t]
\def~{\hphantom{0}}
   \tbl{Variance, squared bias, and mean squared error for mean estimators, Facebook advertising revenue data. Our method outperforms its competitors for a wide range of thresholds $t$.}{%
  \begin{tabular}{lccc}
    Method & Variance \;(s.e.) & Bias$^2$ \;(s.e.) & MSE \;(s.e.) \\[5pt]
    Semiparametric \;(Oracle $t$) & ~12 \;(0.2)~~ & 2.5 \;(0.1) & ~15 \;(0.2)~~\\
    Semiparametric \;($t=0.75$ quantile) & ~28 \;(0.4)~~ & 1.4 \;(0.1) & ~29 \;(0.4)~~\\
    Semiparametric \;($t=0.9$ quantile) & ~40 \;(0.6)~~ & 0.1 \;(0.0) & ~41 \;(0.6)~~\\
    Semiparametric \;(Guillou--Hall) & ~67 \;(0.9)~~ & 2.5 \;(0.3) & ~69
    \;(1.0)~~\\[4pt]
    Winsorized \;(Oracle $t$) & ~72 \;(1.0)~~ & 1.9 \;(0.2) & ~74 \;(1.0)~~\\
    Winsorized \;($k=1$) & ~79 \;(1.2)~~ & 1.3 \;(0.2) & ~80 \;(1.2)~~\\[4pt]
    Pareto Tail \;(Oracle $t$) & ~70 \;(1.0)~~ & 2.0 \;(0.2) & ~72 \;(1.0)~~\\
    Pareto Tail \;(Guillou--Hall) & 642 \;(337.8) & 0.1 \;(0.2) & 643 \;(337.9)
  \end{tabular}}
\label{tab:FB}
\begin{tabnote}
  MSE, mean squared error; s.e., standard error. All numbers multiplied by $10^6$.
\end{tabnote}
\end{table}



\bibliographystyle{biometrika}
\bibliography{references}

\newpage

\begin{appendix}

\section{Proofs}
\label{secDelta}

\subsection{Notation}

Throughout the appendix, we use the notation
$$ \mu_1 = \EE(X \mid X \leq t), \quad \mu_2 = \EE(X \mid X > t), \quad p_2 = \PP(X > t) = 1-F(t).  $$
If $\hmu$ is an estimator for $\mu$, we denote $\MSE(\hmu)=\EE\left\{(\hmu-\mu)^2\right\}$. For an asymptotically normal random variable $Z$, we denote as $\aVar$ the variance of its limiting distribution, and similarly $\aMSE$.
We write $a(n) \asymp b(n)$ to indicate that $a(n)/b(n)$ converges to a finite non-zero limit. If the two are random variables, $a(n) \asymp_p b(n)$ means $a(n)/b(n)$ and $b(n)/a(n)$ are bounded in probability.

\subsection{Proof of Theorem~\ref{theo:var}}

Suppose that $X$ and $Y$ both have regularly varying tails with a common tail index $0 < \gamma < 1$, and that the second-order condition \eqref{eqSecond} holds. Let $t(n)$ be any sequence of thresholds satisfying $t(n)
\rightarrow \infty$ and $t(n) = o(n^\gamma)$. Our goal is to show that, provided that $n^{1/\min(1, \, 2 - 2\gamma)}/N \rightarrow 0$, $\hmu$ is asymptotically normal with
\begin{align}
\label{eqAVar}
n\aVar(\hmu_2) \bigg / \left\{(1 - p_2) \, \Var(X\mid X \leq t) + p_2 (1 - p_2)\, (\mu_2 - \mu_1)^2
+ p_2 \, \frac{\Cov^2(T, X \mid X > t)}{\Var(T \mid  X > t)} \right\} \rightarrow 1.
\end{align}
Our condition on the growth of $t$ is equivalent to requiring $np_2(t)\to\infty$.  In other words, $t$ must grow slowly enough that the number of $X_i$ exceeding the threshold tends to infinity, enabling our estimator of $\mu_2$
to converge.

Given  \eqref{eqSecond}, we can verify that
\begin{equation}
\label{eqMoments}
\PP(X > t) \asymp t^{-1/\gamma}, \quad \EE(X^2 \mid  X < t) \asymp t^{(2\gamma - 1)/\gamma}, \quad \EE(X - t\mid X > t) \asymp t.
\end{equation}
Our mean estimator is $\hmu = (1-\hp_2)\hmu_1 + \hp_2\hmu_2$. Writing $\hmu_1=\mu_1+\ep_1$, $\hmu_2=\mu_2+\ep_2$,
and $\hp_2=p_2+\ep_3$ and rearranging terms, we have
\begin{equation}
\label{eq:errs1}
\hmu - \mu = (1-p_2)\ep_1 + p_2\ep_2 + (\mu_2-\mu_1)\ep_3
+ (\ep_2-\ep_1)\ep_3.
\end{equation}
Equation \eqref{eqMoments} suggests that the first
term has variance
on the order of $n^{-1}t^{(2\gamma-1)/\gamma}$.  Inflating
\eqref{eq:errs1} by this factor and compensating for the order of $p_2$ and
$\mu_2-\mu_1$, we obtain
\begin{align}
\label{eq:errs2}
\sqrt{\frac{n}{t^{{(2\gamma-1)/\gamma}}}}\,\left(\hmu -
  \mu\right)
&= (1-p_2)\, \sqrt{\frac{n}{t^{{(2\gamma-1)/\gamma}}}}\;\ep_1
\;+\; \frac{p_2}{t^{-1/\gamma}}\,
\sqrt{\frac{nt^{-1/\gamma}}{t^2}}\;\ep_2\\ \notag
&\;\;\;\;+\frac{\mu_2-\mu_1}{t}\, \sqrt{nt^{1/\gamma}}\;\ep_3
\;+\; \frac{\ep_2-\ep_1}{t}\, \sqrt{nt^{1/\gamma}}\;\ep_3.
\end{align}
Making the substitutions
\[
Z_1=\sqrt{\frac{n}{t^{{(2\gamma-1)/\gamma}}}}\,\ep_1, \quad
Z_2=\sqrt{\frac{nt^{-1/\gamma}}{t^2}}\,\ep_2, \quad
Z_3=\sqrt{nt^{1/\gamma}}\,\ep_3,
\]
\eqref{eq:errs2} simplifies to
\begin{align}
\label{eq:Zs}
\sqrt{\frac{n}{t^{{(2\gamma-1)/\gamma}}}}\,\left(\hmu -
  \mu\right)
&= (1-p_2)\, Z_1
\;+\; \frac{p_2}{t^{-1/\gamma}}\, Z_2\\ \notag
&\;\;\;\;+\; \frac{\mu_2-\mu_1}{t}\, Z_3
\;+\; \sqrt{\frac{t^{1/\gamma}}{n}}\left(Z_2-t^{-1/\gamma}Z_1\right)\, Z_3.
\end{align}
If we can show that the $Z_i$ converge weakly to independent normal random
variables, then we will have established that the
fourth term of \eqref{eq:Zs} tends in probability to 0, and hence that
the left-hand side is asymptotically normal with variance \eqref{eqAVar}.
Here, the coefficients multiplying $Z_1, \, Z_2, \, Z_3$ converge to
finite limits as $t \rightarrow \infty$, whereas the fourth coefficient converges
to 0 because $t = o(n^\gamma)$ by hypothesis.

Write
\[
Z_1 = \sqrt{nt^{1/\gamma}}t^{-1}\left(\hmu_1-\mu_1\right)
= \sum_{X_i<t} \sqrt{\frac{t^{1/\gamma}}{n}} t^{-1}(X_i-\mu_1).
\]
Since the magnitude of the summand is bounded by
$\sqrt{t^{1/\gamma}/n}\to 0$, the Lindeberg condition is
satisfied and so, by the central limit theorem for triangular arrays and
\eqref{eqMoments}, there is a sequence $s_1$ such that
$s_1 \asymp 1$ and $s_1 Z_1$ is asymptotically standard normal.
Similarly, write
\[
Z_3 = \sqrt{nt^{1/\gamma}}\left(\hp_2-p_2\right) =
\sum_{i=1}^n\sqrt{\frac{t^{1/\gamma}}{n}} \left(\1_{\{X_k>t\}}-p_2\right).
\]
The magnitude of the summand is again bounded by
$\sqrt{{t^{1/\gamma}} / {n}}\to 0$, and so a similar argument applies.

Finally we turn to asymptotic normality of $Z_2$. By our regular variation assumption, $\PP\left\{(X - t)/t \leq x \mid X > t\right\}$
and $\PP\left\{(Y - t)/t \leq x \mid Y > t\right\}$ both converge to Pareto limits both in law and in moments. Thus, assuming that $n = o(N)$, we can use standard exponential family results to show that
$ n_2^{1/2}\left[\heta - \EE\left(\heta\right)\right] $
converges in distribution to a Gaussian random variable with variance $\sigma^2_\eta$,
where $\sigma^2_\eta$ is characterized by
$$ \lim_{n \rightarrow \infty} \PP\left\{\left|\Var_{\heta}(T)^{-1} - \sigma^2_\eta\right| > \varepsilon\right\} = 0 $$
for all $\varepsilon > 0$. Define
$$ J(\heta,t) = \frac{\partial}{\partial \eta} {\EE_{\eta}\left(X/t\right)} \bigg|_{\eta = \heta}.$$
By the delta method, if $J(\heta,t)$ converges in probability to a finite non-zero limit, then
$$ \sqrt{n_2 \, \sigma^2_\eta}  J(\heta,t) \left[\frac{\hmu - \EE\left(\hmu\right)}{t}\right]$$
converges in distribution to a standard Gaussian distribution.
Now, again by standard exponential family theory derivations, we can verify that ${J(\heta,t) = \Cov_{\heta}\left(X, \, T; \, \hG_0\right)},$
where the argument $\hG_0$ emphasizes that we use the empirical background tail as the carrier measure.
Because we know that the the tail law $G_0$ converges to a generalized Pareto distribution as $t$ gets large, we can check that $\Cov_{\heta}(X, \, T; G_0) / \Cov(X, \, T)$ converges in probability to $1$. Thus, if we could replace $\hG_0$ by $G_0$ in the above bound, we would be done.

Thus, we need to show that, provided $N$ is large enough, our use of $\hG_0$ versus $G_0$ has a negligible effect. Based on results for the convergence of heavy-tailed means to stable distributions \citep[e.g.,][]{lepage1981convergence}, we can verify that
$$ \frac{\Cov_{\heta}(X, \, T; \, \hG_0) - \Cov_{\heta}(X, \, T; \, G_0)}{t} = \mathcal{O}_P\left(N_2^{-\min(1 - \gamma, \, 1/2)}\right), $$
where $N_2$ is the number of background tail observations. Thus, provided that
\begin{equation}
\label{eq:n2}
 n_2^{1/2} \big / N_2^{\min(1 - \gamma, 1/2)} \rightarrow 0,
\end{equation}
we find that
\begin{equation}
\label{eq:background}
n_2^{1/2} \frac{\Cov_{\heta}(X, \, T; \, \hG_0) - \Cov_{\heta}(X, \, T; \, G_0)}{t}
\end{equation}
converges in probability to 0, and the desired result holds. Now, because $F$ and $F_0$ both have the same tail index and the threshold sequence $t$ goes to infinity, $n_2/N_2 \asymp n/N$. Moreover, if we inflate $n_2$ and $N_2$ by the same factor, then \eqref{eq:n2} can only become harder to satisfy. Thus, \eqref{eq:background} will also hold whenever
$$ n^{1/2} \big / N^{\min(1 - \gamma, 1/2)} \rightarrow 0. $$

\subsection{Proof of Theorem~\ref{theo:conv}}
\label{secRate}

We start by deriving the risk of Winsorization. Recall that the estimator we call
the Winsorized estimator for the mean is
$ \hmu^{(t)}_{W} = n^{-1}\sum_{i = 1}^n \min(X_i, t)$,
and so by \eqref{eqMoments}
$$ \Var\left(\hmu^{(t)}_{W}\right) \asymp n^{-1}t^{(2\gamma-1)/\gamma}, \; \Bias^2\left(\hmu^{(t)}_{W}\right) \asymp t^{(2\gamma - 2)/\gamma}. $$
To compute the variance estimate, we used the fact that $\EE(X)$ is finite while $\Var(X)$ is infinite, and so
$$ \Var\left(\hmu^{(t)}_{W}\right) \asymp \EE\left\{\left(\hmu^{(t)}_{W}\right)^2\right\}. $$
The MSE can then be minimized at a threshold $t^* \asymp n^\gamma$, giving us an optimal error
$\MSE_{W}^* \asymp n^{2\gamma - 2}$,
as claimed in \eqref{eqRate}.


Next we discuss the risk of our semiparametric method under \eqref{eqSecond}. Because
$$ n^{1/\min(2 - 2\gamma, \, 1)} \big/ N \rightarrow 0, $$
the randomness of the background is negligible just as in the proof of Theorem \ref{theo:var} and so we can effectively treat the background as fixed.
By our delta-method estimate, for $t = o(n^\gamma)$, the asymptotic variance of
our method is
\begin{align*}
\aVar\left(\hmu^{(t)}_{S}\right)
&\sim \frac{\Var(X; X < t)}{n} + \frac{\PP(X > t)}{n}\frac{\Cov^2(T, X \mid  X > t)}{\Var(T \mid  X > t)} \\
& \ \ \ \ + \frac{\left\{\EE(X\mid X > t) - \EE(X\mid X<t)\right\}^2}{n}\PP(X > t)\{1-\PP(X>t)\} \\
&\asymp \frac{t^{(2\gamma-1)/\gamma}}{n} + \frac{t^{-1/\gamma}}{n}\frac{t^2}{1} + \frac{t^2}{n}t^{-1/\gamma} \\
&\asymp \frac{1}{n} t^{(2\gamma-1)/\gamma},
\end{align*}
where on the second line we used moment estimators from
\eqref{eqMoments}. The fact that $\Cov(T, X \mid  X > t) \asymp t$ and
$\Var(T \mid  X > t) \asymp 1$ can be verified by calculus because the scale
parameter $\sigma_0$ grows proportionally to $t$ \citep[][p. 75]{coles2001introduction}.

So far, we have seen that given any shared threshold sequence, the
variance of our semiparametric estimator decays at the same rate as
that of Winsorization. Now, we show that under~\eqref{eqSecond}, the bias of our method decays faster than that of
Winsorization, which enables to use smaller thresholds and
achieve better risks.

For convenience, let $\tX = X - t$ and $\tY = Y - t$, conditional
respectively on $X$ and $Y$ exceeding the threshold $t$. We consider the exponential tilts
$ dG_\eta = e^{\eta T - \psi(\eta)} dG_0$ with $\eta \in \mathbb{R}$,
and fit the distribution $G$ of the $\tX$ with
$G_{\eta^*}$, where $\eta^*$ is the population MLE for $\eta$.
If we write $\mu(\eta)$ for the mean of $dG_\eta$, we find
that the bias of our semiparametric estimator is given by
$$ \aBias\left(\hmu^{(t)}_{S}\right) = \PP(X > t) \, \left[\mu(\eta^*) - \EE\left(\tX\right) + \oo\left\{\frac{t}{n\PP(X > t)}\right\}\right], $$
where $\aBias$ is the bias of the center of limiting normal distribution of $\hmu^{(t)}_{S}$.
Here, the main term is due to model misspecification arising from the
fact that $\tY$ and $\tX$ are only converging to the generalized Pareto distribution.  The
remainder, which will turn out not to affect the decay rate of the asymptotic mean squared error, is due to higher-order curvature effects (i.e., the second-order term in the delta-method expansion).

We begin by establishing a tail bound. As a consequence of \eqref{eqSecond}, we find that
$$ \frac{1 - G(tx)}{1 - G(t)} = x^{-1/\gamma} \ \left\{1 + D_Xt^{-\beta_X}(x^{-\beta_X} - 1) + o(t^{-\beta_X})\right\}. $$
This implies that
\begin{align*}
\frac{1 - G\left\{t\left(1+x\right)\right\}}{1-G(t)}
= \left(1 + x\right)^{-1/\gamma} \ \left[ 1 + D_X t^{-\beta_X} \left\{\left(1 + x\right)^{-\beta_X} - 1\right\}+ o(t^{-\beta_X})\right],
\end{align*}
or, in terms of the statistic $T = x/(\kappa + x)$,
\begin{align}
\label{eq:tail_bound}
&\PP\left\{T(\tX - t) > \tau\right\} = \left(1 + \frac{\tau}{1 -
    \tau}\frac{\kappa}{t}\right)^{-1/\gamma}
     \left[ 1 + D_X t^{-\beta_X} \left\{\left(1 + \frac{\tau}{1 -
    \tau}\frac{\kappa}{t}\right)^{-\beta_X} - 1\right\} + o(t^{-\beta_X})\right]
\end{align}
for $0\leq \tau < 1$. A similar expression holds for $\PP\{T(\tY - t) >
\tau\}$. Notice that the $o(t^{-\beta_X})$ term is bounded in $x$ as $x$
gets large; this means that we can use \eqref{eq:tail_bound} to
establish the convergence of moments. Now, recalling that $\kappa \sim t$, we can use the tail bound \eqref{eq:tail_bound} to establish many useful relations.

First, because $\tX$ and $\tY$ share the same tail bound with possibly different constants $D$ and $\beta$, we see that
\begin{equation}
\label{eq:T_diff}
\EE\left\{T\left(\tX - t\right)\right\} - \EE\left\{T\left(\tY - t\right)\right\} \asymp t^{-\beta_{\min}},
\end{equation}
where $\beta_{\min}$ is the smaller of the two second-order constants.
Second, recall that in an exponential family, the MLE $\eta^*$ is defined by the relation
$$ \EE\left\{T\left(\tX - t\right)\right\} = \EE\left\{e^{\eta^* T(\tY - t) - \psi(\eta^*)}T(\tY - t)\right\}. $$
We have already seen that
$$ \frac{d\EE\left\{e^{\eta T(\tY - t) - \psi(\eta)}T(\tY - t)\right\}}{d\eta} \bigg|_{\eta = 0} = \Var\left\{T\left(\tY - t\right)\right\}, $$
which by \eqref{eq:tail_bound} converges to a finite non-zero limit as the threshold $t$ goes
to infinity. Thus, because of \eqref{eq:T_diff}, we see that $\eta^{*} \asymp t^{-\beta_{\min}}$.

Finally, doing some calculus, we can use \eqref{eq:tail_bound} to show that
$$ \frac{\EE\left(\tX\right) - \EE\left(\tY\right)}{\EE\left\{T\left(\tX - t\right)\right\} - \EE\left\{T\left(\tY - t\right)\right\}} =
\frac{(1+\gamma)(1 + \gamma + \gamma\beta_{\min})}{(1-\gamma)(1 - \gamma - \gamma\beta_{\min})} \, t + o(t). $$
We are now ready to bound the bias of our method. By the same arguments as in the proof of Theorem~\ref{theo:var}, we find that
$$\frac{1}{t} \frac{d\mu(\eta)/d\eta}{d\EE_{G_{\eta}}(T)/d\eta} \bigg|_{\eta = 0}
= \frac{1}{t}\frac{\Cov\left(\tY, \, T\right)}{\Var\left\{T\left(\tY - t\right)\right\}}
= \frac{(1 + \gamma)(1 + 2\gamma)}{(1 - \gamma)\gamma} + o(1),
$$
and that the second-order term also has a finite limit. Thus,
\begin{align*}
\frac{1}{t}\left\{\mu(\eta^*) - \EE\left(\tX\right)\right\}
&= \frac{1}{t}\left[\mu(\eta^*) - \EE\left(\tY\right) - \left\{\EE\left(\tX\right) - \EE\left(\tY\right)\right\}\right] \\
&= \left\{\frac{(1 + \gamma)(1 + 2\gamma)}{(1 - \gamma)\gamma} + o(1)\right\} \,  \left[\eta^* + \oo\left\{(\eta^*)^2\right\}\right] \\
&\ \ \ \ - \left\{\frac{(1+\gamma)(1 + \gamma + \gamma\beta_{\min})}{(1-\gamma)(1 - \gamma - \gamma\beta_{\min})} + o(1)\right\} \, \left[\EE\{T(\tX - t)\} - \EE\{T(\tY - t)\}\right] \\
&\asymp t^{ - \beta_{\min}},
\end{align*}
and so finally
$$\aBias\left(\hmu^{(t)}_{S}\right)
\asymp \PP(X > t) \, \left[t^{1 - \beta_{\min}} + \oo\left\{\frac{t}{\PP(X > t)n}\right\}\right]
\asymp t^{(\gamma - \gamma\beta_{\min} - 1)/\gamma} + \oo\left(\frac{1}{t^{-1/\gamma}n}t^{(\gamma - 1)/\gamma}\right)
$$
Putting all the pieces together, we get
$$  \aMSE\left(\hmu^{(t)}_{S}\right) \asymp \frac{1}{n} t^{(2\gamma-1)/\gamma} + t^{{(2\gamma - 2\gamma\beta_{\min} - 2)/\gamma}} + \oo\left(\frac{1}{t^{-1/\gamma}n}t^{(2\gamma - \gamma\beta_{\min} - 2)/\gamma}\right), $$
where the asymptotic mean-squared error $\aMSE$ describes the limiting normal distribution of $\hmu^{(t)}_{S}$. This is optimized with $t^*(n) \asymp n^{\gamma/(1 + 2\gamma\beta_{\min})}$, which leads to
\begin{equation}
\label{eq:aMSE}
\aMSE^*_{S} \asymp n^{(2\gamma - 2 - 2\gamma\beta_{\min})/(1 + 2\gamma\beta_{\min})} \ \left\{1 + \oo\left(n^{-\gamma\beta_{\min}/(1 + 2\gamma\beta_{\min})}\right)\right\}.
\end{equation}
For any $\beta > 0$, our optimal threshold sequence satisfies the relation $t^*(n) = o(n^\gamma)$ which we assumed at the beginning.

\subsection{Proof of Corollary \ref{coro:GH}}

As shown above \eqref{eq:aMSE}, achieving the optimal rate of convergence from Theorem \ref{theo:conv}
only requires that as $n$ grows, our threshold $t_n$ grows as
$t^*(n) \asymp n^{\gamma/(1 + 2\gamma\beta)}$.
\citet{guillou2001diagnostic} frame their problem as choosing the number of order statistics $\hkGH$ with which to compute the Hill estimator. A choice of $\hkGH$ immediately implies a threshold, namely the $\hkGH$ largest observation
$\htGH = X_{n - \hkGH + 1, \, n}. $
Under \eqref{eqSecond}, they show that
$ \hkGH / n^{2\gamma\beta/(1 + 2\gamma\beta)} \asymp_p 1. $
Now, we also know that $\PP(X > t) \asymp t^{-1/\gamma}$, which implies that
$$ \htGH \asymp_p \left({n^{2\gamma\beta/(1 + 2\gamma\beta)}} \Big/ {n}\right)^{-\gamma} = n^{\gamma/(1 + 2\gamma\beta)}. $$

\section{Additional simulation results}

\begin{figure}
    \centering
      \includegraphics[width=.75\textwidth]{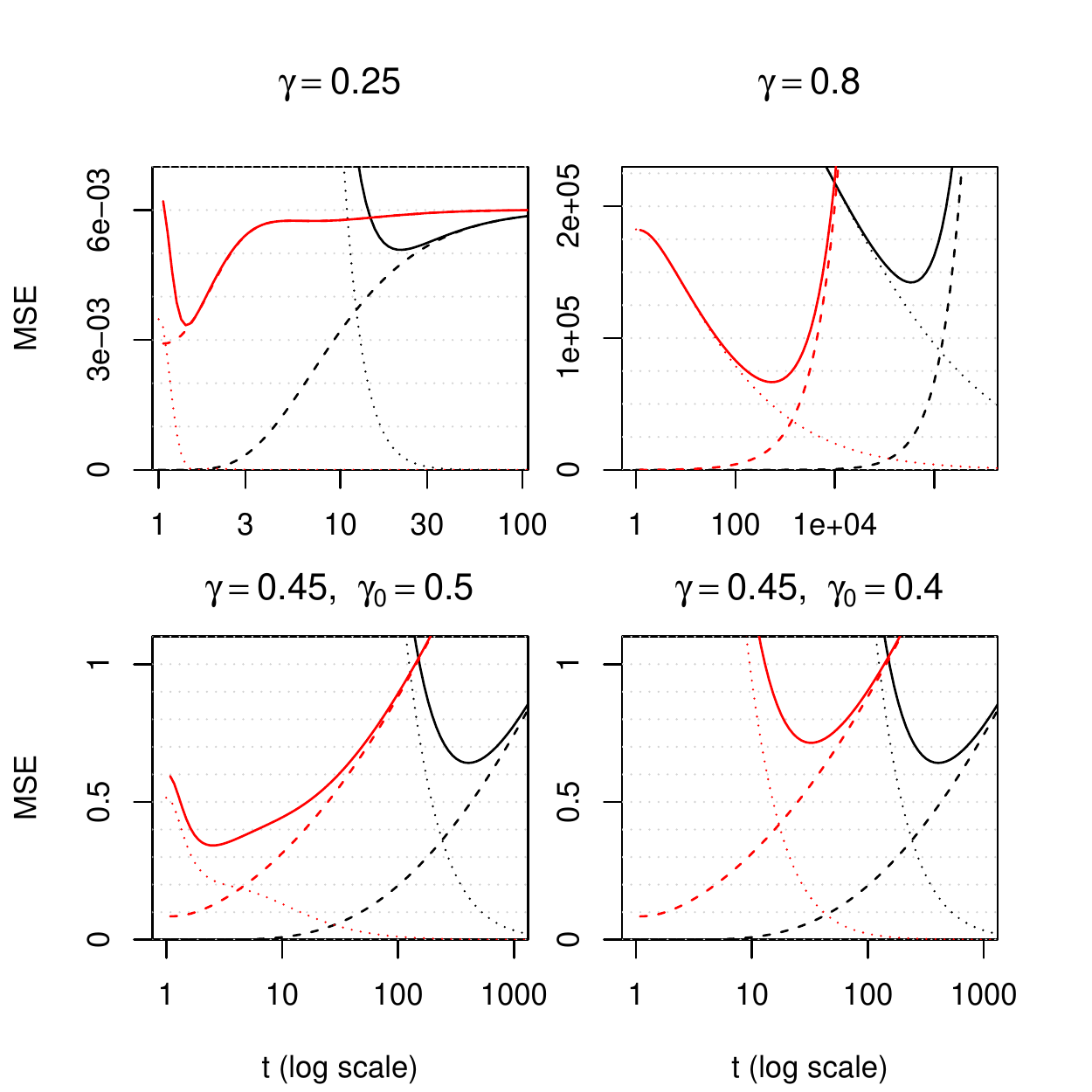}
      \caption{Squared bias (dotted lines), variance (dashed), and mean squared error (MSE, solid) for mean estimation using Winsorization (black) and our method (red), in four variations on the simulation of \S 4.2. In the top panels, the tail index $\gamma$ is the same $X$ and $Y$. In the lower panels, $Y$ has tail index $\gamma_0\neq \gamma$, so our model is misspecified.}
\label{fig:simVarious}
\end{figure}

 Figure~\ref{fig:simVarious} plots the same bias--variance tradeoff curves for four variations on the log-gamma simulation. We compare the methods on data with other tail indexes, $\gamma=0.25$ and $\gamma=0.8$; in both cases, our method comfortably outperforms Winsorization and the parametric Pareto tail method. Table~\ref{tab:simResultsOtherGamma} shows the squared bias, variance, and mean squared error for each method using both an optimal fixed threshold and a threshold chosen adaptively, as in Table~\ref{tab:simResults}.

We also stress-test our method under violations of our major assumption that the two samples have the same tail index. Without changing anything else about the simulation of \S\ref{secSimStudy}, we give the background data a different tail index $\gamma_0\neq \gamma$. In two different scenarios, we set $\gamma_0=0.4$ and $0.5$, leaving $\gamma=0.45$. It appears that $\gamma_0=0.4$ exacerbates our method's bias while $\gamma=0.5$ introduces some offsetting bias, counterintuitively improving our method's performance. Table~\ref{tab:simResultsMisspec} shows simulation results in the same format. These simulations suggest that, while violation of our assumptions can affect our method's performance, it is robust to small differences in the two tail indices.

\begin{table}
\def~{\hphantom{0}}
   \tbl{Bias, variance, and mean squared error for mean estimators, for variations on the log-gamma simulation. For both $\gamma=0.25$ and $\gamma=0.8$, our method comfortably outperforms its competitors.
For $\gamma=0.8$, the Pareto tail method had mean squared error approximately $4\times 10^6$ and $6\times 10^7$, respectively for oracle $t$ and Guillou--Hall thresholds.}{%
  \begin{tabular}{l|ccc|ccc}
    \multicolumn{1}{c}{}
    &\multicolumn{3}{c}{$\gamma = 0.25$}
    &\multicolumn{3}{c}{$\gamma = 0.8$}\\[5pt]
    Method & Var \;(s.e.) & Bias$^2$ \;(s.e.) & MSE \;(s.e.)
    & Var \;(s.e.) & Bias$^2$ \;(s.e.) & MSE \;(s.e.)
    \\[5pt]
    Semiparametric \;(Oracle $t$)
    & 35 \;(0) & 2 \;(0) & 36 \;(1)
    & 10 \;(1) & ~96 \;(1) & 106 \;(1)\\
    Semiparametric \;(Guillou--Hall)
    & 47 \;(1) & 3 \;(0) & 50 \;(1)
    & ~3 \;(0) & 136 \;(0) & 140 \;(0)\\[4pt]
    Winsorized \;(Oracle $t$)
    & 48 \;(1) & 3 \;(0) & 51 \;(1)
    & 25 \;(1) & 116 \;(1) & 141 \;(1)\\
    Winsorized \;($k=1$)
    & 56 \;(1) & 1 \;(0) & 57 \;(1)
    & 23 \;(6) & 161 \;(1) & 185 \;(5)\\ [4pt]
    Pareto Tail \;(Oracle $t$)
    & 53 \;(1) & 0 \;(0) & 54 \;(1)
    & $-$ & $-$ & $-$ \\
    Pareto Tail \;(Guillou--Hall)
    & 63 \;(1) & 0 \;(0) & 63 \;(1)
    & $-$ & $-$ & $-$ \\
  \end{tabular}}
\label{tab:simResultsOtherGamma}
\begin{tabnote}
  Var, variance; MSE, mean squared error; s.e., standard error. All numbers for $\gamma=0.25$ multiplied by 10,000; all numbers for $\gamma=0.8$ multiplied by 0.001.
\end{tabnote}
\end{table}

\begin{table}
\def~{\hphantom{0}}
   \tbl{Bias, variance, and mean squared error for mean estimators, for misspecified versions of the log-gamma simulation. Here, $\gamma$ is the tail index for the population of interest, and $\gamma_0$ is the index for the background population.}{%
  \begin{tabular}{l|ccc|ccc}
    \multicolumn{1}{c}{}
    &\multicolumn{3}{c}{$\gamma = 0.45,\;\gamma_0 = 0.5$}
    &\multicolumn{3}{c}{$\gamma = 0.45,\;\gamma_0 = 0.4$}\\[5pt]
    Method & Var \;(s.e.) & Bias$^2$ \;(s.e.) & MSE \;(s.e.)
    & Var \;(s.e.) & Bias$^2$ \;(s.e.) & MSE \;(s.e.)
    \\[5pt]
    Semiparametric \;(Oracle $t$)
    & ~16 \;(0)~~ & ~4 \;(0) & ~20 \;(0)~~
    & ~42 \;(1)~~ & 15 \;(0) & ~57 \;(1)~~\\
    Semiparametric \;(Guillou--Hall)
    & ~37 \;(1)~~ & ~0 \;(0) & ~37 \;(1)~~
    & ~32 \;(1)~~ & 82 \;(1) & 114 \;(1)~~\\[4pt]
    Winsorized \;(Oracle $t$)
    & ~50 \;(2)~~ & 14 \;(2) & ~64 \;(3)~~
    & ~50 \;(2)~~ & 14 \;(2) & ~64 \;(3)~~\\
    Winsorized \;($k=1$)
    & ~78 \;(5)~~ & 12 \;(2) & ~90 \;(5)~~
    & ~78 \;(5)~~ & 12 \;(2) & ~90 \;(5)~~\\ [4pt]
    Pareto Tail \;(Oracle $t$)
    & ~64 \;(3)~~ & ~9 \;(2) & ~74 \;(3)~~
    & ~64 \;(3)~~ & ~9 \;(2) & ~74 \;(3)~~\\
    Pareto Tail \;(Guillou--Hall)
    & 472 \;(269) & 17 \;(6) & 488 \;(273)
    & 472 \;(269) & 17 \;(6) & 488 \;(273)\\[7pt]
  \end{tabular}}
\label{tab:simResultsMisspec}
\begin{tabnote}
  Var, variance; MSE, mean squared error; s.e., standard error. All numbers multiplied by $100$.
\end{tabnote}
\end{table}

\end{appendix}

\end{document}